\documentclass[10pt, conference]{IEEEtran}

\AtBeginDocument{%
  \providecommand\BibTeX{{%
    \normalfont B\kern-0.5em{\scshape i\kern-0.25em b}\kern-0.8em\TeX}}}
    
\usepackage{soul}
\usepackage[switch]{lineno}

\usepackage{balance}
\usepackage{booktabs}
\usepackage{graphicx}
\usepackage{multirow}
\usepackage[table, dvipsnames]{xcolor}
\usepackage{array}
\usepackage{ifthen}
\usepackage{amssymb}
\usepackage{hyperref}
\usepackage{listings}
\usepackage[normalem]{ulem}
\newboolean{showcomments}
\setboolean{showcomments}{true} 
\ifthenelse{\boolean{showcomments}}
  {\newcommand{\nb}[2]{
    \fcolorbox{gray}{yellow}{\bfseries\sffamily\scriptsize#1}
    {\sf\small$\blacktriangleright$\textit{#2}$\blacktriangleleft$}
   }
   
  }
  {\newcommand{\nb}[2]{}
   
  }

\colorlet{LightRed}{red!50!}
\colorlet{LightGray}{gray!50!}

\definecolor{codegreen}{rgb}{0,0.6,0}
\definecolor{codegray}{rgb}{0.5,0.5,0.5}
\definecolor{codepurple}{rgb}{0.58,0,0.82}
\definecolor{backcolour}{rgb}{0.95,0.95,0.92}

\lstdefinestyle{mystyle}{
    backgroundcolor=\color{backcolour},   
    commentstyle=\color{codegreen},
    keywordstyle=\color{magenta},
    numberstyle=\tiny\color{codegray},
    stringstyle=\color{codepurple},
    basicstyle=\ttfamily\footnotesize,
    breakatwhitespace=false,         
    breaklines=true,                 
    captionpos=b,                    
    keepspaces=true,                 
    showspaces=false,                
    showstringspaces=false,
    showtabs=false,                  
    tabsize=2
}

\newcommand{\datadoi}{%
  \begingroup\normalfont
  \smash{\href{https://doi.org/10.5281/zenodo.4457108}{\includegraphics[height=1.3\fontcharht\font`\B,trim=0 3 0 0]{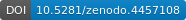}}}%
  \endgroup
}

\begin{document}


\title{Representation of Developer Expertise in Open Source Software}
\author{\IEEEauthorblockN{Tapajit Dey}
\IEEEauthorblockA{
The University of Tennessee\\
Knoxville, TN, USA\\
Email: tdey2@vols.utk.edu}
\and 
\IEEEauthorblockN{Andrey Karnauch}
\IEEEauthorblockA{
The University of Tennessee\\
Knoxville, TN, USA\\
Email: akarnauc@vols.utk.edu}
\and
\IEEEauthorblockN{Audris Mockus}
\IEEEauthorblockA{
The University of Tennessee\\
Knoxville, TN, USA\\
Email: audris@utk.edu}}

\maketitle
\thispagestyle{plain}
\pagestyle{plain}


\begin{abstract}
Background: Accurate representation of developer expertise has
always been an important research problem. While a number of studies
proposed novel methods of representing expertise within individual
projects, these methods are difficult to apply at an ecosystem
level. However, with the focus of software development shifting from
monolithic to modular, a method of representing  developers'
expertise in the context of the entire OSS development becomes
necessary when, for example, a project tries to find new maintainers 
and look for developers with relevant skills.
Aim: We aim to address this knowledge gap by proposing and constructing the 
\textit{Skill Space} where each API, developer, and project is represented and
postulate how the topology of this space should reflect what developers 
know (and projects need).
Method: we use the World of Code infrastructure to extract the
complete set of APIs in the files changed by open source
developers and, based on that data, employ Doc2Vec embeddings for 
vector representations of APIs, developers, and projects. We then
evaluate if these embeddings reflect the postulated topology of the
\textit{Skill Space} by predicting what new APIs/projects developers
use/join, and whether or not their pull requests get accepted.  We also check how the
developers' representations in the \textit{Skill Space} align with their
self-reported API expertise.  Result: Our results suggest that the
proposed embeddings in the \textit{Skill Space} appear to satisfy the
postulated topology and we hope that such representations may aid in the
construction of signals that increase trust (and efficiency) of open source
ecosystems at large and may aid investigations of other phenomena
related to developer proficiency and learning.
\end{abstract}




\section{Introduction}\label{s:intro}
The number of projects and developers involved with open source
software has reached staggering heights, e.g. GitHub reported 
that over 10 million new developers joined and 
over 44 million new projects were created in 2019
alone~\footnote{\url{https://octoverse.github.com/}}. While many of
these developers or projects are based on individual effort, further
statistics, such as over 87 million pull requests being merged and 20
million issues being closed in the past year on GitHub alone, demonstrate
that open source development is a highly collaborative effort.  

The key premise of open source software is not only
to share the code, but, more importantly, to enable
contributions from the community~\cite{cathedralandbazaar,MFH02}.
However, despite improved tools and
practices enabled by social coding platforms such as GitHub, it is
not always easy to get contributions accepted, and, as many studies have
shown, repeated interactions between the maintainers and
contributors are necessary to establish trust and increase the chances
of pull request acceptance or issue
resolution~\cite{gousios2014exploratory,zhu2016effectiveness,guo2010characterizing,dey2020effect,TsaySocialAndTechnical,dey2020pull}.
However, this method of building reputation and trust by repeated interactions 
does not scale very well, and, with a growing number of developers and an increasing
number of projects their code may depend on~\cite{ssc}, 
other means of establishing trust are becoming necessary.
Previous work~\cite{DabbishHiringSignals,TsaySocialAndTechnical} has shown
that both \textit{technical and social aspects of a developer's reputation}
can play an important role in building the \textit{trust} between themselves
and other developers. While social aspects, such as previous
collaboration ~\cite{HahnPriorCollaboration}, can greatly increase
the trust between two developers, these aspects are not broadly
applicable as they enhance trust within an already-established
developer circle. For a developer looking to contribute for the
first time to a project outside of their social circle, the technical aspect
of their reputation, often referred to as \textit{expertise}, may
serve as an important \textit{source of trust} for other developers when evaluating a
developer as a potential team member or collaborator~\cite{lucassen2011factual}. 

We, therefore, concentrate on gauging the \textit{relevant} 
expertise of a developer based on their previous development activities.
Such measure, if it could be obtained,
might partially substitute for the traditionally laborious
reputation building process as a developer transitions from a peripheral
participation in a project to a contributor role~\cite{jergensen2011onion},
and could potentially increase the efficiency of the open source 
development as a whole. However, previous attempts at measuring developer 
expertise either focus on very detailed views, e.g. counting ``experience atoms''
associated with changes made by a developer on a specific source
code file~\cite{MH02}, or, at the least granular level, counting the
volume, frequency, and breadth~\cite{montandon2019identifying,kam19}
of a developer's overall activities.  Unfortunately, the former
approach can not be applied for developers who have never
participated in a specific project, while the latter does not
account for the specific experience of a developer beyond the aggregated
activity traces and projects they've worked on. Aggregates of
developer's contributions by programming language was previously
proposed by Amreen et al.~\cite{kam19}, however, experience in a particular
language does not immediately confer experience in the variety of
libraries or frameworks 
in that language which specific applications might rely on. 
However, this measure of \textit{domain expertise}, or expertise measured by the
fluency of using specific APIs, is something that may be of greater
concern to projects~\cite{montandon2019identifying} than a potential
contributor's overall skill in a language.

In this work, we try to measure and evaluate such specific domain expertise by
defining what we refer to as \textit{Skill Space}, that can be
applied to developers, projects, and individual programming languages or APIs as well.
In other words, \hl{\textit{Skill Space} provides a vector
representation for individual developers, projects, programming languages, or APIs,
with the topology of the resulting representations (\textit{skill vectors}) reflecting the conceptual and practical (API-related) relationships among these four
entities.}  

To operationalize this \textit{Skill Space} we use the World of Code (WoC)~\cite{woc19,ma2020world} data
that contains APIs extracted from changes to source code files (discussed further in Section~\ref{ss:apis}) in 
17 programming languages. We employ Doc2Vec~\cite{doc2vec} text embedding that uses as input the 
dependencies (APIs) of a file modified in each change made by a developer to produce 
the \textit{Skill Space} representation for
individual APIs, developers, projects, and languages. The topology in this space is defined by 
the alignment (cosine similarity) between vectors representing any pair
of developers, projects, APIs, developers and APIs,  developers and
projects, and  projects and APIs. 

Compared to similar other methods (see Section~\ref{s:relwork}), using \textit{Skill Space} offers the following practical advantages: a) ability to compare the developers, projects, and the APIs in the same space; b) a more faithful representation of expertise due the completeness of the training data (entire OSS); c) cross-language comparison; d) up-to-date representation of expertise based on the latest version of World of Code (WoC) dataset.

Our key contributions consist of a) conceptualization of 
developer skill/expertise that transcends individual project boundaries making
it
specific enough to determine its relevance in a novel context; b) postulating the desirable
topology of the resulting \textit{Skill Space}; c) proposing Doc2Vec embedding method
for operationalizing the \textit{Skill Space}; and d) an empirical evaluation of the proposed
topology for this operationalization. 
A replication package for this paper is made available at \datadoi~\cite{dey_tapajit_2021_4457108}.

In the rest of the paper, we start by describing the specific research problems in
Section~\ref{s:research}. The related works are described in
Section~\ref{s:relwork}. We describe our methodology in Section~\ref{s:method}, 
and the evaluation results for our proposed embedding for the proposed \textit{Skill Space} is
described in Section~\ref{s:result}. Details of the replication package we shared is described in Section~\ref{s:repl}. We describe the limitations to our study in
Section~\ref{s:limitation}, the planned extension of the proposed technique in 
Section~\ref{s:future}, and conclude the paper in Section~\ref{s:conclusion}.

\section{Research Problem}\label{s:research}
Our \textbf{aim} in this paper is to \textit{define a feasible representation
  of a developer's expertise in specific focus areas of
  software development} by gauging their fluency with different APIs. Such
medium-granularity representation of developer expertise might serve as a way 
to get a better understanding of
developer skill, help recommender systems that suggest APIs, projects,
or contributors, or to increase the trust between external
contributors and maintainers of a project.  To achieve these goals we
define the concept of \textit{Skill Space} and we propose the desirable
properties and an operationalization of this concept.  We
quantify \textit{Skill Space} based on the World of Code
(WoC)~\cite{woc19} data that contains information on the APIs
extracted from changes to source code files (discussed further in
Section~\ref{ss:apis}) in 17 programming languages.

\subsection{Postulated properties of \textit{Skill Space}}\label{ss:properties}
The critical feature of our concept of \textit{Skill Space} is the
ability to make direct comparisons among three entities: developers,
projects, and APIs. The simplest way to accomplish that is to
represent each entity as a vector in a linear space. Once such
representation is accomplished, for it to be meaningful it needs to
satisfy several simple properties: \textit{First}, we expect that the skill vectors of APIs
representing similar skills will be close to each other; \textit{Second},
a developer's skill vector should be similar to the representation of
the APIs they use most frequently; \textit{Third}, a project's skill vector should be
similar to the representations of the APIs used in these
projects; \textit{Finally}, we expect the developer representations to be
aligned with their subjective perceptions of their API mastery. 

Apart from these four fundamental properties, for \textit{Skill Space} 
to be useful in practice, we expect a few additional properties
to be satisfied: \textit{First}, in order to predict API usage, we expect
that the new APIs a developer will use in the future should have representations
more similar to the representations of APIs they have used in
the past compared to randomly selected APIs; \textit{Second}, we expect
that new APIs added in projects should also follow a similar
pattern; \textit{Third}, we expect that developers will be more likely
to join new projects that have representations similar to themselves
in the \textit{Skill Space}. We also expect other manifestations of
``good'' \textit{Skill Spaces} in terms of outcomes of developer
work, e.g.
the closeness between the skill vector of a developer who submitted 
a pull request (PR) to a project and that of the target project should 
have a significant impact on the PR acceptance probability.

\textit{Skill spaces} satisfying these properties can obviously be
of practical and theoretical use, hence our objective in this paper
is to construct such a \textit{Skill Space} and to evaluate if it
satisfies these desirable properties. 

\subsection{Operationalization of \textit{Skill Space}}
To produce the representations in the \textit{Skill Space} we follow
previous successful approaches such as degree-of-knowledge model~\cite{FM10} and
experience atom~\cite{MH02} that take the uncontroversial position that 
developer's skill increases as they complete and repeat tasks requiring a specific skill. In the context of software engineering, that
involves making changes to the source code. Since we are trying to
capture the experience of using programming APIs, we capture the
APIs that a modified source code file depends upon. We further
discuss the pros and cons of this choice and potential alternatives in
Section~\ref{s:limitation}. Since many of the software source code files are
an approximation of software modules~\cite{parnas-criteria}, the
collection of the APIs a file depends upon should represent a specific
use case of the functionality instantiated by the file and should, thus,
provide implicit dependencies between the APIs utilized in that
file. The entirety of all source code, thus, should embody all
realized relationships among APIs. Once these implicit
relationships among APIs based on changes to the source code are
captured, the representation of a developer in the \textit{skill
  space} could simply be derived from the changes they have
made, the representation of a project through changes made in
that project, and the representation of a programming language
through all changes involving that language.

A naive representation of each change would simply be a
high-dimensional vector\footnote{We counted over 100 million
  distinct import/use/package/etc. statements in the programming
  languages from WoC version R} that represents each of the distinct
APIs extracted from over 4 billion changes to the source code files
of the languages under consideration. However, such representation
of APIs in the \textit{Skill Space} is not very effective or
practical, and techniques from text analysis~\cite{word2vec} may be
used to reduce the dimensionality of this vector. The key underlying
assumption of text analysis techniques is that words in a natural
language are used in certain combinations to express certain ideas
or thoughts. The unsupervised approaches where the relationships are
learned directly from the corpus of text assume that the words
within a document have to be related and represent some underlying
idea expressed by that document. For larger documents sliding window
techniques are often used to restrict the length of text where
these assumed relationships among words pertain to the same idea.
Similarly, we assume that a combination of APIs used in a software
module would also reflect some aspects of the functionality
implemented in that module. The number of APIs in a single file tends
to be quite low as we find in Table~\ref{t:stats}, so there is no
need for sliding windows when representing the API. However, text 
analysis methods need a large
corpus of natural language text to extract the semantics from word
combinations. We, similarly, expect that the \textit{Skill Space}
representation would require a very
large corpus of software modules to represent these distinct
functionalities (and the associated skill of developers who
implemented it). In this paper, we use Doc2Vec~\cite{doc2vec} text
embedding approach to produce the \textit{Skill Space}
representation not just for individual developers, but also for
individual APIs, projects, and even languages. As a result, the
proposed \textit{Skill Space} representation can be used to
calculate a direct measure of alignment between any pair of
developers, projects, APIs, developers and APIs, developers and
projects, and projects and APIs.

\subsection{Evaluation criteria}\label{ss:eval-cri}
A conceptual definition also needs practical utility, therefore, to
evaluate the suitability of our proposed \textit{Skill Space}
representation, we investigate several practical scenarios
where developer expertise and trust might come into
play, and we expect that a closer alignment between developers and
APIs or projects in the \textit{Skill Space} will increase the
likelihood of a positive outcome in these events.  Specifically, we
pose the desirable properties of the \textit{Skill Space} (outlined in Section~\ref{ss:properties}) as hypotheses which we evaluate to determine if
the proposed representation of a developer's specific expertise in the
\textit{Skill Space} might be useful in practice by evaluating the
following topological properties of the \textit{Skill Space}:
\begin{itemize}
    \item [H1:] A developer is more likely to choose new APIs that
      are more closely aligned\footnote{Since we use cosine similarity to measure the closeness between
entities, the word ``alignment'' is a better choice than a more
conventional ``distance.''} with themselves.
    \item [H2:] A developer is more likely to join
    new projects that are more closely aligned to themselves.
    \item [H3:] A project is more likely to accept contributions
      from developers who are more aligned with the project.
    \item [H4:] Developers better
      aligned with the project's will have
      better odds to have their pull requests accepted.
    \item [H5:] A developer's self-reported API skills are closely
      aligned to their own representation in \textit{Skill Space}.
\end{itemize}

\section{Related Work}\label{s:relwork}
In this section, we present an overview of the historic efforts to measure developer
expertise and outline the role of word embeddings in the software
engineering literature to clarify the existing gaps we try to address
with our work.
\vspace{-.05in}  
\subsection{Developer Expertise}
The fascination with developer expertise and its variation began
in the early days of software
development~\cite{basili79,AlGa,cu81,Boehm-81}. Early work
was primarily motivated by the need for software project cost
estimation and focused on various ways to measure the size of
software by adjusting lines of code for different languages or
attempting to design ways to have a language-independent measure of
software size~\cite{Behrens-83}. The later works embraced the
idea that beyond language, each software project requires long and
arduous work by a developer to comprehend its internal
complexities~\cite{MM10}. This suggested that developer expertise is
project and file specific with approaches such as Expertise Browser
assuming that each change to a source code file represents an
experience atom~\cite{MH02}, whereby a developer changing code is
forced to understand the files' internal design and, perhaps, impart of
their own design through implementing that change. However, these early measures of lines of code written and file-specific experience atoms pertain to expertise within a specific project. They do not provide a general enough profile of developer expertise that can be transferred among software projects.

Contemporary social coding platforms (e.g. GitHub) provide a variety
of indicators of developer activity (the timeline of
commits) and their social status (followers). This has sparked a variety of research into how developer traces and developer profiles can provide insight into a developer's expertise. These studies include qualitative approaches, such as the one by Marlow et. al.~\cite{marlow2013}, who showed that your developer profile on GitHub can help other developers gauge your general coding ability and project-relevant skills, but only at a more general level. Similarly, Singer et. al.~\cite{singer2013} interviewed developers and employers to observe how they utilize developer profiles to gauge the quality of a potential new hire. The results showed that profile sites with a ``skills'' word-cloud representing the technologies (languages, frameworks, etc.) a developer claimed to be familiar with proved to be the most helpful assessment of a developer's expertise. These works indicate that more specific measures, such as language-specific technologies and frameworks, help others gauge the relevant expertise of developers in open source.

There have also been several attempts to automate the process of identifying developer expertise through social coding platforms, e.g. CVExplorer~\cite{CVExplorer} is a tool created to expose developer expertise using a word-cloud of all
relevant technologies, frameworks, and general skills by parsing their commit messages and README files. SCSMiner~\cite{wan2018scsminer} is another tool created to help identify experts on GitHub based on an arbitrary input query. The authors also obtain expertise attributes by parsing README files of projects a developer has contributed to, but they extend this by creating a generative probabilistic expert ranking model to rank developers based on certain skills or expertise one might be looking for. Lastly, Hauff et. al.~\cite{hauff15jobs} attempt to match developers with job advertisements based on a developer's expertise by extracting relevant terms from README files and mapping them to the same vector space as job advertisements, and ranking all developer profiles based on the cosine similarity they share with the job advertisements. Cosine similarity has been used in similar contexts in a number of earlier studies (e.g. ~\cite{dey2016analysis}) and was also used for evaluating the performances of the Doc2Vec and Word2Vec techniques~\cite{lau2016empirical}. While all of these approaches are a similar step in the same direction as us, they provide a weaker link between developers and their technologies than desired by utilizing README files as the main source of developer expertise, while we extract language-specific APIs from files a developer has modified. Furthermore, along with measuring a developer's similarity to the technologies they use as attempted in previous work, we also aim to use the APIs to measure the similarity between developers, projects, developers and projects, and projects and APIs.


We also motivate our work through some more recent studies. Montandon et. al.~\cite{montandon2019identifying} present an approach to determine experts for three JavaScript libraries. The authors
identify developers who have made changes to projects that depend
on these libraries and conduct a survey with 575 developers to
obtain their self-reported expertise. Using these survey results as validation, the authors argue that their clustering approach is feasible and can be used to identify relevant experts. However, they also present the shortcomings of using basic GitHub profile features for machine learning classifiers to predict expertise in software libraries. We utilize the survey dataset provided by the authors for our own evaluation and also attempt to better predict developer expertise in software libraries, an area in which the authors achieved poor performance.

The more recent Import2Vec~\cite{theeten2019import2vec} paper produces
embeddings for each imported package. The authors do such embeddings for
JavaScript, Python, and Java, and provide some qualitative evidence
suggesting that these embeddings of APIs accurately reflect
different functionality profiles by providing a number of examples
where the similar APIs also appear to implement similar
functionalities.

Unfortunately, none of the proposed approaches are suitable for
directly comparing developers and projects, as neither developers nor projects
are accurately represented in the same vector space as the API
embeddings. It is, therefore, not clear how Import2Vec embeddings can
be used to represent developers' domain expertise nor if such
profiles would accurately reflect developer
proficiency. Furthermore, the Import2Vec approach can not be
applied in a cross-language context. Our proposed approach tries to address this gap by constructing a
\textit{Skill Space} representation that, on one hand, may transcend the specific
programming languages, and on the other hand, may identify a
meaningful representation that can be matched with skill sets of other developers
or projects.

\subsection{Vector Embedding in Software Engineering}
Vector embeddings have been used in software engineering for various
tasks, e.g. using natural language associated with coding to determine
sentiment~\cite{biswas2019exploring}, using writing style in commit
messages to determine developer
identity~\cite{AZBZM19}, or improve requirements
traceability~\cite{guo2017semantically}. In these cases the natural
language techniques do not need to be modified substantially as the
underlying data represents natural language.

Even more techniques have been applied to model programming language
source using text analysis techniques. For example, these approaches
can improve Interactive Development Environments (IDEs) by
performing next token prediction~\cite{allamanis2018survey}, suggesting better class 
names~\cite{allamanis2015suggesting}, or even automatic
patching~\cite{marginean2019sapfix}.  In a recent paper, Alon et
al.~\cite{alon2019code2vec} proposed a method for representing snippets 
of code as continuous distributed vectors (code embeddings).

The attempt to provide a common embedding space for natural language
and code was proposed by Ye et al.~\cite{ye2016word} by training the natural
language models on the API documentation and the
applications that use these APIs.

Unlike these approaches, we focus on training the models on the
APIs used in files that undergo a code change. While we do not go to
the level of a specific function used in the API, we treat each
import/use statement as an indication of the specific functionality
provided by the corresponding package.
As noted above, the best natural language analysis techniques
typically exploit the order of the words in a text document (such as
commit messages, requirements, or documentation). The programming
language modeling techniques also rely heavily on the specific
sequence that is necessary to do an accurate prediction of the next
token, for example.  
In contrast, our work looks at embedding package imports within source code files,
where the order of import statements may not be important. Thus,
the existing techniques that attempt to model the order of the tokens
need to be modified to fit our purpose.

\section{Methodology}\label{s:method}
To represent our entities in the \textit{Skill Space} we need a
very large corpus of software and we turn to World of Code (WoC)
due to its size, coverage, data quality, and the ability to obtain
desirable subsamples as described below.
\vspace{-5pt}
\subsection{Data Source: World of Code}
WoC is a prototype of an updatable and
expandable infrastructure, aimed at supporting research and tools that rely on
version control data from open source projects that use Git. 
It stores large and rapidly growing amounts of data that approximates the entire FLOSS
ecosystem, and provides capabilities to efficiently extract
and analyze the data at that scale.
In addition to storing objects from all
git repositories, WoC also provides relationships among them.  
The primary focus of WoC is on the types of analyses that require global reach
across FLOSS projects, so it is the most appropriate choice for answering the
research questions we presented here. 

WoC data is versioned, with the latest version labeled as R, 
containing 7.9 billion blobs, 2 billion commits, 8.3 billion trees, 17.3 
million tags, 123 million projects (distinct repositories), and 42 million
distinct author IDs. This version of WoC data was collected during
March, 2020. 

As is often the case with datasets of this size, certain data
cleaning steps are critical for obtaining meaningful results. Conveniently,
in addition to providing access to the raw
data, WoC offers advanced data augmentation capabilities.  Two such
techniques were used in this study for data preprocessing: fork
resolution (deforking) and developer identity resolution, since our
\textit{Skill Space} representation considers the relationship
among projects, developers, and their API usage.
Accurately representing all three of these entities is, therefore,
necessary.
\subsubsection{Project Clones: Fork Resolution}
Git is a distributed version control system that, inherently,  
makes it easy to clone or fork Git projects. This, however, creates a
unique data cleaning problem for WoC, which has over 116 million
projects, many of which are clones or forks of another project.
This
poses several problems for our expertise analysis. One such problem
is that a developer who contributes to a highly-cloned project will
have their commits appear in the remaining cloned projects as well,
e.g. if a developer contributes to one project using the
\texttt{flask} module in Python and 10 other people clone this
project and make little to no changes, the developer would be
attributed with having worked with \texttt{flask} on 11 different
projects, rather than just one. 

To address this, we use the dataset published in~\cite{forks20},
which applies the Louvain community detection algorithm to a massive
graph consisting of links between commits and projects in WoC
(because two projects are highly unlikely to share the same exact
commit unless they are clones). We leverage that work to combine
commits from the forked projects and ensure that we do not count the
same project-related information multiple times due to these
forks/clones. 
\subsubsection{Identifying a Developer: Identity Resolution}
The WoC dataset contains the author ID 
for each git commit, which would, ideally,
correspond to a single developer, and could be used to
aggregate all commits associated with the author ID and perform our
expertise analysis. However, this is seldom the case as the author ID
is obtained from the git configuration file residing on
the developer's laptop/desktop/server where they use git. The author ID tags,
therefore, often differ between commits made on different computers
used by a developer. As a result, many developers have multiple
author IDs (with some that they might not even be aware of) in WoC
collection that, collectively, need to represent the same developer.

To address this, we have used a dataset shared by Fry et
al.~\cite{fry2020idres} that resolves the 38 million author identities
in WoC version Q by creating blocks of potentially related author
IDs (e.g. IDs that share the same email, unique first/last name) and
then predicting which IDs actually belong to the same developer
using a machine learning model. The approach identified over 14
million author IDs belonging to at least one other author ID. From
this set, around 5.5 million developers were identified, with a
median of two author IDs per developer. When performing the
expertise analysis described in this paper, we identify each
developer using the new associations created by the identity
resolution approach. This allows us to create a much more accurate
representation of each developer's API usage and expertise and helps
us avoid comparing two author IDs that are in fact the same
developer. 
\subsection{API Extraction}\label{ss:apis}
To obtain developer API usage, we utilize the language mappings
inside WoC. These mappings contain APIs extracted from changes to
source code files in C, 
C\#, Java, FORTRAN, Go, JavaScript, Python, R, Rust, Scala, Perl,
Ruby, Dart, Kotlin, TypeScript, and Julia languages, as well as source code present in
Jupyter (iPython) Notebooks~\footnote{https://jupyter.org/}. The
mappings are created by first obtaining all files in WoC with
extensions used by each of the languages listed previously. For each
language, the WoC file-to-blob\footnote{\url{https://git-scm.com/book/en/v2/Git-Internals-Git-Objects}} map is used to obtain all blobs
associated with language-specific files.
The content of the resulting blobs is then parsed for 
import statements depending on the syntax of each language
(e.g. \texttt{\#include} in C, \texttt{import} in 
Java/Python, \texttt{use} in Perl, the dependencies in the
package.json file for npm, and so forth).

Each of these blobs (versions of the source code) is further mapped
to the commit(s) that produced it and projects that have that
commit. Timestamps, authors, and projects of these commits are then associated
with the blob as well as with the APIs parsed from that blob
resulting in the following tuple (programming language, repository,
 timestamp, author id, timestamp, API1, ...).
We use deforking and author aliasing described above to transform
repository into deforked project ID and author id into aliased
developer id. The timestamp allows us to perform time-based
prediction in some of our models as discussed in
Section~\ref{ss:eval}.

Thus, the final mapping and data used by some of the 
models is a compressed file of
entries containing:\\ \texttt{project;timestamp;developer;API1;API2;...},
where each entry represents all modules/APIs included in the file
that the developer added to the project at the instance in
time. There is a unique set of entries for each language listed
earlier, and they are stored in separate compressed files. While this
mapping serves as the base data for most of our analysis, there are
several intermediate steps that require a transformation of the
provided mapping as well.
\subsection{Summaries of API usage}
\begin{table*}[tbp]
\caption{Summary of data retrieved from WoC-version.R per Language}
\label{t:stats}
\resizebox{\linewidth}{!}{
\begin{tabular}{|r|>{\raggedleft\arraybackslash}p{3cm}|r|r|>{\raggedleft\arraybackslash}p{1.5cm}|>{\raggedleft\arraybackslash}p{3cm}|>{\raggedleft\arraybackslash}p{4cm}|}\toprule
  \textbf{Language}&\textbf{Delta (Changed blobs)}&\textbf{Authors}&\textbf{Projects}&\textbf{Distinct APIs}& \textbf{Fraction of deltas (changed blobs) with 30 or fewer APIs} & \textbf{Max no. of APIs in one delta (changed blob)} \\\hline
FORTRAN & 1,628,760 & 24,898 & 15,623 & 59,349 & 0.98 & 106 \\
Julia & 1,297,134 & 18,666 & 35,723 & 104,725 & 0.99 & 108 \\
R & 6,822,662 & 361,754 & 516,678 & 85,255 & 0.998 & 117 \\
iPython & 12,160,775 & 793,261 & 1,154,120 & 687,085 & 0.99 & 1,158 \\
Perl & 18,780,774 & 480,615 & 547,115 & 58,942 & 0.999 & 109 \\
Rust & 13,599,452 & 95,712 & 148,327 & 818,686 & 0.99 & 118 \\
Dart & 7,036,000 & 116,317 & 164,360 & 467,863 & 0.99 & 165 \\
Kotlin & 28,129,485 & 281,469 & 429,071 & 6,233,673 & 0.96 & 1,096 \\
TypeScript & 239,416,852 & 1,605,563 & 2,253,291 & 7,324,019 & 0.99 & 1,013 \\
C\# & 220,871,444 & 2,092,316 & 3,092,761 & 6,648,357 & 0.997 & 150 \\
Go & 123,432,323 & 490,967 & 662,355 & 245,102 & 0.995 & 1,207 \\
Scala & 36,361,141 & 176,414 & 210,175 & 3,571,593 & 0.99 & 1,288 \\
Ruby & 74,618,824 & 1,222,886 & 2,343,825 & 669,297 & 0.997 & 1,002 \\
JavaScript & 55,609,812 & 3,362,191 & 7,347,050 & 1,105,918 & 0.67 & 10,014 \\
Python & 612,708,423 & 4,795,735 & 6,820,899 & 17,227,676 & 0.99 & 1,001 \\
C/C++ & 1,780,602,124 & 3,656,965 & 4,704,446 & 2,553,521 & 0.99 & 1,007 \\
Java & 1,106,084,606 & 5,063,200 & 7,512,800 & 85,079,403 & 0.92 & 1,004 \\\bottomrule
\end{tabular}
}
\vspace{-.1in}  
\end{table*}
The previous subsection describes the procedures used to obtain the
data from WoC (version R) that captures for each modification to
the source code the programming language, the timestamp, the developer,
the project, and the list of ``import'' statements.


Table~\ref{t:stats} shows the number of deltas (changed blobs) associated
with each language as well as the number of distinct authors and
projects involved. 
The largest number of delta by far involve C
and C$++$ (we do not distinguish between the two), followed by Java
and Python. The relatively low number of JavaScript
delta relates to the way dependencies are specified in JavaScript
projects where a single file (\texttt{Package.json}) is used to specify the
dependencies while in C, Java, or Python, every source code file
needs to include its dependencies explicitly.

Notably, Java language dominates in terms of the number of unique
APIs, presumably because the APIs in Java can be specified using global
namespace, while for other languages they are defined by the package
managers or within the source code files (like .h files in C/C$++$) that may share
the same name but be otherwise unrelated (see Section~\ref{s:limitation}).

As noted above, the total number of distinct APIs we observe is
far higher than the number of words in a natural language putting
computational strains on the text analysis methods designed to
deal with many orders of magnitude smaller dictionaries.
Moreover, the order of the APIs in
source code files is not important, hence we need to apply methods
that do not attempt to model the sequences. While some early text analysis
methods, such as LSI, work strictly on the bag of words (BOW) and
are immune from this problem. Others, such as continuous bag of
words (CBOW), try to predict words within a certain window size. The wider
the window, the more complicated and time consuming it is to fit
these models. To investigate what window sizes might be appropriate,
we investigate the distribution of the number of distinct APIs
within a single delta (a modification by a single commit to one
source code file). 

Table~\ref{t:stats} shows the fraction of delta for
each language where the number of distinct APIs is less than 30
and also shows the maximum number of APIs. Again, JavaScript
is an outlier here since a single file (package.json) defines APIs for the entire
project. We chose to consider the window size of 30 or
less for the CBOW models since it captures most of the deltas for all languages.
The deltas with huge numbers of
APIs used may indicate unusual cases or outliers that may not bring
much information to which APIs are used together and it is not unreasonable 
to exclude those from consideration.


The total number of delta and the number of distinct APIs pose
serious computational challenges if we want to fit the complete
dataset obtained from WoC with 4.3B delta and over 100M distinct APIs
not counting the number of distinct projects and authors. 
We, therefore, fit several smaller datasets by filtering the data to a more
manageable size.

First, for the multi-language model, we focus on
developers that made between 100 and 25K commits partially to
exclude the bot activities and partly to consider ordinary
but productive developers, since by the premises of our proposed hypotheses, 
we're trying to focus on developers who have a good amount of contributions in
social-coding platforms, since our assumption is that they will use new APIs,
contribute to multiple projects, and will submit a number of pull requests.
This filter reduces the total number of delta
down to 1.2B. For language specific models we are dealing with much smaller
datasets, but we can decrease that size even further by 
randomly sampling projects or developers. We used these smaller samples to debug 
the techniques and to find the parameters for the \textit{Skill Space} embeddings 
that produce feasible results before running the
computation on the entire model.

\subsection{Vector Embedding}
Since the total number of possible APIs that can be used by a
developer or a project across different languages is extremely large
and the naive embedding, representing API usage as a component, of
over a 100M-dimensional vector is not practical, we reduce
the dimensionality of the \textit{Skill Space}. We chose to employ Doc2Vec
embedding method since it is capable of embedding not only the APIs
themselves but developers and projects at the same time. It is also one of
the most efficient embeddings to compute: an important consideration
given the large data corpus we handle.

Word2Vec,~\cite{word2vec} is a highly computationally efficient
algorithm used to create a numerical representation for a word using
a continuous bag of words or skipgram (two distinct algorithms). The
primary assumption of Word2Vec is that only words that are close
together in a document are semantically related. In our
context, that assumption doesn't hold, because there is no semantic
order for the APIs used by a developer or a project. We 
address this potential problem by using the continuous bag of words
algorithm with a wide window of 30 words. Since the number of
APIs associated with a single blob rarely exceeds 30 as shown in
Table~\ref{t:stats}, the algorithm in practice predicts one API of
a blob using all remaining APIs.

Doc2Vec is an extension of Word2Vec, where in addition to word (API)
embeddings, the model also produces the embeddings for an arbitrary
set of tags associated with a group of APIs, as is the case when an
author, a project, and a language is associated with the set of APIs
extracted from each change of every file. The continuous bag of
words analog in Doc2Vec corresponds to obtaining doc-vectors by
training a neural network on the synthetic task of predicting a
word based on an average of both context word-vectors and the
full document's doc-vector. We used the Gensim framework
for evaluation due to its high performance.


\subsection{Evaluation strategies}\label{ss:eval}
The evaluation strategy involves fitting a Doc2Vec model on past
data, where each document represents the APIs encountered in a
single delta and the document tags represent the
language, the project, and the developer. The resulting model thus
creates vectors for each API, for each developer, each project, and
each language. We then obtain new APIs a developer uses during the
testing period, the new projects the developer joins, and the new
developers who join a project during the testing period. The
alignment to these factual APIs/projects/developers are then
compared with randomly chosen sets of APIs/projects/developers of
the same size.

We chose the dates so that we have a fairly short testing period
starting from February, 2019. All changes prior to that date were
used to fit the model and the activities past that date to check the
predictions. We used these dates for predicting new APIs, developers
joining new projects, and projects accepting new contributors.

For PR acceptance and self-reported expertise, we fitted models based
on data prior to Feb 14, 2018 and tested on activities after that
time in order to have a sufficient number of accepted or rejected PRs
during the testing period for most developers.
To conduct the study of pull request acceptance, we sourced the pull request
dataset~\cite{tapajit_dey_2020_3858046} used by Dey and Mockus~\cite{dey2020effect} for verifying our hypothesis
and studying the effects of technical and social factors on PR acceptance. The
dataset contained information on 470,925 PRs from 3349 popular NPM packages 
and 79,128 GitHub users who created those. We filtered this dataset to only include
developers who made between 100 and 25,000 commits, similar to what we did for
testing earlier hypotheses. In addition, we removed small projects that didn't 
have any API calls. After filtering, we were left with 150,173 PRs 
made by 14,784 developers for 1860 GitHub projects.


Then, as in the other cases,  we proceeded to obtain embeddings for the
developers and projects using past data and then model the
acceptance rate during the future PR activity using the binomial
regression with the independent variable representing the 
alignment of the developer and project vectors where the PRs have
been submitted to together with the predictors used by
~\cite{dey2020effect}. We once again use February, 2018 to separate
training and test data. 

Finally, we use a previously reported survey~\cite{montandon2019identifying} 
of JavaScript
developers to compare how aligned each surveyed developer is to the
the API in which developers were reported to be proficient. Since
the survey did not include APIs where developers reported being not
proficient, we randomly chose ten other APIs under the assumption
that they might not be equally proficient in these 10 randomly
chosen APIs. As in other comparisons, we report the difference in
alignment between the self-reported expert APIs and the randomly
chosen APIs. To make the \textit{Skill Space} representations
commensurate with developer self-reported expertise, we only use the
data close to the time when the survey was conducted (also February, 2018).  

Given the very large vocabularies for the APIs, we chose a relatively
high-dimensional vector of 200 for \textit{Skill Space}, to make
sure there is enough flexibility to represent the extremely
large number of potential skills. We excluded APIs that occur in fewer than
five deltas to increase computational efficiency and, also, avoid
highly uncertain embeddings. As discussed above, we chose a window
size of 30 to ensure that the order of APIs in the delta
does not matter.  Finally, we chose the negative sampling parameter to
be 20. It tends to speed up the convergence by creating synthetic
samples (API combinations) that do not exist in the data and
penalizes the model if it produces a good fit for such ``negative''
samples. All of these parameters were chosen after extensive
experimentation fitting the models on manageable-size datasets. 
\section{Results}\label{s:result}
\subsection{Qualitative Evaluation of Skill Space Embeddings}
For a qualitative evaluation of our proposed embedding, we decided to observe which
APIs
are reported as similar to others in the same language, 
and also which APIs
provide similar functionality across different languages.
For the Python package ``pandas'', we observed that the APIs reported to
be most similar are indeed the ones that are most frequently used with it, primarily
for data manipulation/ data visualization/ machine learning applications.
\begin{lstlisting}[language=Python]
>>>mod.most_similar('pandas')
>>>[('matplotlib.pyplot', 0.8), ('numpy', 0.8), ('seaborn', 0.78) ]
\end{lstlisting}
We can also do some arithmetic with the resulting vectors by
asking what are packages the most similar to Python ``pandas'' package in R
language:
\begin{lstlisting}[language=Python]
>>> mod.wv.similar_by_vector(-mod.docvecs['PY']  + mod.docvecs['R'] + mod.wv.get_vector('pandas'))
>>> [('data.table', 0.83), ('dplyr', 0.82) ]
\end{lstlisting}
As we see, the most popular data frame (after which ``pandas'' was modeled)
packages are most similar. Also, only R packages appear in the most similar
list even though we start from the python package and move in the
direction of R. 
\subsection{Examining H1: New APIs used by developers are closely aligned to themselves in the
    \textit{Skill Space}}
\begin{table}[!tbp]
\centering
\caption{Summary of per-Language Results of t-test showing the difference of alignments between a developer's representation in the \textit{Skill Space} and the APIs they used in future vs. random APIs they didn't use (in the same language). p-Values $<$1e-200 are shown as 0.}
\label{t:api-res}
\resizebox{\linewidth}{!}{%
\begin{tabular}{@{}l>{\raggedleft\arraybackslash}p{2.5cm}>{\raggedleft\arraybackslash}p{2.5cm}r@{}}
\toprule
Language & Estimated Difference in Means & 95\% Confidence   Interval & p-Value \\ \midrule
Dart & 0.41 & 0.39  - 0.43 & 3.12e-92 \\
Julia & 0.21 & 0.15  - 0.27 & 8.57e-05 \\
R & 0.14 & 0.09  - 0.20 & 1.46e-06 \\
iPython & 0.20 & 0.18  - 0.22 & 6.68e-65 \\
Perl & 0.05 & 0.03  - 0.06 & 2.85e-13 \\
Rust & 0.21 & 0.20  - 0.22 & 2.01e-151 \\
Kotlin & 0.21 & 0.20  - 0.22 & 1.09e-139 \\
TypeScript & 0.23 & 0.22  - 0.24 & 0 \\
C\# & 0.25 & 0.23  - 0.26 & 6.16e-137 \\
Go & 0.15 & 0.14  - 0.15 & 0 \\
Scala & 0.20 & 0.19  - 0.22 & 8.45e-89 \\
Ruby & 0.17 & 0.16  - 0.18 & 3.80e-188 \\
Java & 0.13 & 0.12  - 0.13 & 0 \\
C/C++ & 0.13 & 0.13  - 0.13 & 0 \\
Python & 0.12 & 0.12 - 0.12 & 0 \\
JavaScript & 0.10 & 0.10  - 0.10 & 0 \\ 
\rowcolor{LightRed}
FORTRAN  & -0.11 & -0.73 - 0.51 &  0.268\\ \bottomrule
\end{tabular}%
}
\end{table}    

We follow the process outlined in 
Section~\ref{ss:eval} to get the alignment between embeddings of each
developer, created by the APIs they used during the training period, and 
the new APIs used in the testing period and a set of random APIs \textit{in the same
language} that they did not use. We did the calculation separately for each 
language to get a clearer understanding of the performance of our proposed
\textit{Skill Space} embeddings at that level. 

We were unable to fit model for the entire corpus (it would
have taken several months on a fast multi-processor
server). Instead we sampled 36K projects 
that contain 1.2B delta by 690K authors in all 17
languages. The amount of data for each language is similar to that
in the entire corpus.


The paired t-test results in Table~\ref{t:api-res} show that the APIs
used in the future were indeed more closely aligned as compared to
random APIs they didn't use. The amount of data for the FORTRAN
language in the sample was too small to get a statistically
significant difference.

\subsection{Examining H2: A developer is more likely to join
    a new project that is more closely aligned to them in the
    \textit{Skill Space}}

Here we try to validate the expectation that the new projects a
developer will join (make an accepted contribution to) would 
be more closely aligned with the developer's \textit{Skill Vector} than a randomly selected project. 

As described in Section~\ref{ss:eval}, 
we calculated the alignment between embeddings of each developer and the projects
they contributed to and a set of random other projects \textit{in the same language}
that they did not contribute to, and measured if there is any significant 
difference between them using t-test. We found there is indeed a significant
difference (p-value $<$ 2.2e-16) with a difference between the estimated means of
the cosine similarity of $0.017$ and 95\% confidence interval of $[0.013, 0.021]$.
This supports our hypothesis that there is a similarity between the
developers vectors and vectors of the projects they contribute to in future.

\subsection{Examining H3: A project is more likely to accept contributions from developers who are aligned to the project in the \textit{Skill Space}}

One of the potential \textit{Skill Space} applications is increasing
trust. New contributors who have \textit{Skill Vectors} aligned to a
project's \textit{Skill Vectors} should be more likely to have their
contributions accepted all other factors being equal. Their skill (if it
exists) should manifest itself in the technical aspects of the PR
and, therefore, might be recognized by the maintainers of that
project. Once again, we constructed skill vectors for the developers
who contributed to a project, measured the alignment between them
and the skill vectors of the corresponding projects, and compared
them with the alignment between skill vectors of a project and the
skill vectors of randomly chosen developers who did not contribute
to that project. The differences between the alignments were found to
be significant using t-test, with p-value $<$ 2.2e-16, an estimated
difference of means between the alignments being 0.141, and a 95\%
confidence interval of $[0.126, 0.156]$.

\subsection{Examining H4: A developer whose \textit{Skill Space} is aligned more closely to the project's \textit{Skill Space} will be more likely to have their pull requests accepted}

\begin{table}[t]
\centering
\caption{Result of Logistic Regression model predicting PR acceptance.\textit{Cosine Similarity between Developer and Project} is the variable we introduced in this study (highlighted in gray). Other variables are adopted from~\cite{dey2020effect}. The non-significant variable is highlighted in red, binary variables are in Blue}
\label{t:pr-res}
\resizebox{\linewidth}{!}{%
\begin{tabular}{@{}p{3.5cm}rr@{}}
\toprule
\textbf{Predictor} & \textbf{Coefficient} $\pm$ \textbf{Std. Error} & \textbf{p-Value}   \\ \midrule
(Intercept) & 0.654 $\pm$ 0.093 & 2.24e-12 \\
\rowcolor{LightGray}
\textit{Cosine Similarity between Developer and Project} & \textit{0.396 $\pm$ 0.084} & \textit{2.10e-06} \\
creator\_submitted & -0.120 $\pm$ 0.009 & $< 2e-16$ \\
creator\_accepted & 0.874 $\pm$ 0.033 & $< 2e-16$ \\
repo\_submitted & -0.026 $\pm$ 0.005 & 1.62e-06 \\
repo\_accepted & 2.864 $\pm$ 0.056 & $< 2e-16$ \\
\textcolor{blue}{dependency:1} & -0.212 $\pm$ 0.021 & $< 2e-16$ \\
age & -0.221 $\pm$ 0.004 & $< 2e-16$ \\
comments & -0.173 $\pm$ 0.013 & $< 2e-16$ \\
review\_comments & 0.342 $\pm$ 0.011 & $< 2e-16$ \\
commits & -0.360 $\pm$ 0.015 & $< 2e-16$ \\
additions & -0.015 $\pm$ 0.008 & 0.05 \\
deletions & -0.035 $\pm$ 0.006 & $< 2e-16$ \\
changed\_files & -0.151 $\pm$ 0.016 & $< 2e-16$ \\
\textcolor{blue}{contain\_issue\_fix:1} & 0.123 $\pm$ 0.020 & 1.89e-09 \\
\textcolor{blue}{user\_accepted\_repo:1} & 1.326 $\pm$ 0.027 & $< 2e-16$ \\
creator\_total\_commits & 0.086 $\pm$ 0.009 & $< 2e-16$ \\
creator\_total\_projects & 0.015 $\pm$ 0.007 & 0.029 \\ 
\rowcolor{LightRed}
\textcolor{blue}{contain\_test\_code:1} & -0.418 $\pm$ 0.324 & 0.197 \\\bottomrule
\end{tabular}%
}
\vspace{-10pt}
\end{table}

To more directly evaluate the previous hypothesis, here we restrict
our attention to Pull Requests (formal external contributions) where
we can see not only the cases when the contribution was accepted as
above, but also cases where the contribution was made but not
accepted. As previously, we hypothesize
the developers' alignment with projects in \textit{Skill Space}
should have a significant impact on PR
acceptance probability, with a better alignment being associated
with a higher chance of acceptance.

We used a regression model for this analysis, as mentioned in Section \ref{ss:eval}.
The result of the Logistic Regression model is presented in Table~\ref{t:pr-res}, which shows that the alignment between developers and projects remains a significant variable even after accounting for the other social and technical factors described in~\cite{dey2020effect}, i.e. this variable describes a factor which is not captured by other technical and social factors. We also notice that the coefficient for this variable is positive, i.e. the closer a developer's alignment is to a project, the higher the chance of their PR being accepted, which validates our proposed hypothesis. We checked the Variance Inflation Factors for these variables and found the values to be less than 2.5 in all cases, signifying that there is no multicollinearity effect.
The variable `contain\_test\_code' was found to be insignificant, similar to~\cite{dey2020effect}. However, the variable `deletions' was found to be insignificant in \cite{dey2020effect} but it's significant here, which could be because we're only focusing on a subset of the data used in that study.

\vspace{-5pt}
\subsection{H5: A developer's self-reported API skills are closely aligned to  themselves} 

The final question we pose is whether the representations in
\textit{Skill Space} align with developer's self-reported opinions about their own
expertise related to a specific technology.

We obtained data from the replication package of~\cite{montandon2019identifying}
that surveys a sample of GitHub users to create a ground truth for
self-reported developer expertise in the studied libraries. In this
survey, the participants declared their expertise (on a scale from 1
to 5) for three JavaScript libraries: \textit{mongodb}, \textit{react},
and \textit{socketio}.

Similarly to previous experiments, we obtain \textit{skill space}
representations for survey participants and the three APIs.
We investigate if the skill space similarity can be explained
by the self-reported score by fitting a linear regression model and
find that the self-reported score explains increases in alignment to each API 
as self-reported expertise score increases. The result of the linear regression 
model is shown in Table~\ref{t:DSLR}(A).


Finally, we try to model the self-reported score
using the amount of activity (commits) as reported
in~\cite{montandon2019identifying} and adding the \textit{Skill Space}
similarity. Again, we find that the increase in skill alignment
has a statistically significant positive relationship with the
self-reported score even after adjusting for the direct measure of
experience based on the number of commits. The result of the model is 
shown in Table~\ref{t:DSLR}(B).

\begin{table}[t]
\vspace{-10pt}
\caption{Result of Linear Regression models: (a) explaining Developer-API Alignment ($R^2$ value: 0.90); (b) explaining self-reported Skill Score  ($R^2$ value: 0.92)}
\label{t:DSLR}
\centering
\textbf{\large{(A)}}\\
\centering
\resizebox{\linewidth}{!}{%
\begin{tabular}{|llr|}
\toprule
Predictors & Estimate $\pm$ Std. Err. & p-Value \\ \midrule
API:mongodb & 0.249 $\pm$ 0.013 & $<$ 2e-16 \\
API:react & 0.307 $\pm$ 0.011 & $<$ 2e-16 \\
API:socketio & 0.422$\pm$ 0.012 & $<$ 2e-16 \\
log(No. of Commits) & 0.000$ \pm$ 0.001 & 0.9 \\
Self-Reported Score & 0.014$\pm$ 0.003 & 1.8e-6 \\ \bottomrule
\end{tabular}%
}\\
\vspace{10pt}
\centering
\textbf{\large{(B)}}\\
\resizebox{\linewidth}{!}{%
\begin{tabular}{|llr|}
\toprule
Predictors & Estimate $\pm$ Std. Err. & p-Value \\ \midrule
API:mongodb & 2.5 $\pm$ 0.10 & $<$ 2e-16 \\
API:react & 2.9 $\pm$ 0.08 & $<$ 2e-16 \\
API:socketio & 1.9 $\pm$ 0.12 &$<$ 2e-16 \\
log(No. of Commits) & 1.1$\pm$0.012 & $<$ 2e-16 \\
Developer-API Alignment & 0.98 $\pm$ 0.21 & 1.81e-6 \\ \bottomrule
\end{tabular}%
}
\hfill
\vspace{-10pt}
\end{table}

In summary, we find that the proposed \textit{Skill Space} embedding based on
Doc2Vec models of the APIs in files changed by a developer has 
a strong and statistically significant relationship with the
self-reported developer expertise. Furthermore, even after adjusting
for the less granular measure of experience (number of commits), we
still see that \textit{Skill Space} representation has a strong
explanatory power.  


\section{Replication Package}\label{s:repl}

The replication package for this paper  is made available through Zenodo under CC 4.0 license at \datadoi~\cite{dey_tapajit_2021_4457108}. The data we share include the input data (processed), with the details of the APIs in each blob modified by OSS developers who made between 100 and 25,000 commits, all the scripts used by us for the evaluation, and the steps for replicating the results presented in the paper (in the README file).
Although we do not share it as a tool/package, which would be difficult to run without access to the World of Code dataset (we are working on extending the publicly available capability of the World of Code dataset, which would make such a tool practical in near future), we share the input data, so that researchers can fit their own models and experiment with the dataset. We provide a detailed account of the steps we took and share the scripts we used so that researchers can replicate our findings. We also share the pre-trained Doc2Vec models, so that researchers can use them for their applications without having to re-train the model.

\section{Limitations}\label{s:limitation}

It is important to note the primary objective behind introducing the concept of   \textit{Skill Space}:
the ability to compare developers, projects, languages, and APIs with the
ultimate goal of better measuring developer skills and at
facilitating ways to make open source software development more
effective by creating signals about the developers' expertise that
is more general than the modification of individual files, but more
specific than their volume of overall activity.

The objective of this work is to conceptualize \textit{Skill Space},
to list some of its properties, and to demonstrate that it is
possible to construct it on a very large corpus of programming
languages and APIs.  As such, we focus on demonstrating the feasibility and novel applications enabled by the proposed measure rather than trying to compare our method with existing ones since  existing developer expertise measures are not suitable for directly comparing developers, projects, and the APIs used by/in them.

Our results, consequently, have to be interpreted with care.  First,
our definition of developer skill is constructive and practical. We
are only concerned that it reflects postulated measures of
performance and has some agreement with developers' subjective
perceptions. Further work is needed to ascertain if it satisfies any
additional properties or is suitable for non-constructive
definitions of skill.

Specifically, the definition of \textit{Skill Space} we chose is based on
API usage, but the skill embeddings can be conducted for other types
of skills as well.

We validate the proposed \textit{Skill Space} by checking if
it would satisfy the intuitive properties the \textit{Skill Space} should
exhibit, but there may be additional properties we do not consider
(and the proposed \textit{Skill Space} does not satisfy). For example, our
primary concern in this work is to capture the aspects of developer
expertise related to the APIs they use and we are not concerned with
other types of expertise, such as their proficiency to do good
design, architecture, testing, and so forth, or with their ability
to communicate with other developers.

The particular mechanism of what it means to use an API may be refined. We only
consider if the version of the file modified by a developer has
certain import statements, but do not verify that the API is
actually exercised in the file, and we also do not check if the
developer made a change to the part of the code that exercises a
specific subset of the API used in the file. Moreover, it can be
argued that just because a developer uses some API in a file doesn't
mean that they are expert in using that API since code snippets are
often copied and pasted from different sources. However, our
assumption is that a developer should have a basic familiarity with
the APIs used in the files they modify, at least more than a random
other API they have never been associated with, and, as noted by 
Lucassen and Schraagen~\cite{lucassen2011factual}, \textit{``domain
familiarity can be seen as a weaker form of domain expertise.''}

Since our aim is to capture the profile of expertise as a
trust-building support and we attempt to create such measures that
equally apply to individual APIs, projects, and developers, there are no
golden datasets that could be created to evaluate the objectivity of
all such measures. Specifically, there is no convincing test everyone
would agree upon that a developer is a good fit for a project. As such,
we can evaluate the goodness of the measures we propose through
several indirect means e.g., can a specific developer be trusted
when they make a contribution if there has been no prior interaction
between the developer and maintainer? As we noted above, different
languages have different conventions in which APIs are declared and
these differences may play a role or need to be taken into account
in order to improve upon the proposed implementation of the skill
space. 

There are a few other shortcomings associated with our approach,
e.g. our method of measuring expertise can't be applied to complete
newcomers, since they likely have worked with very few APIs, and
their representation in the \textit{Skill Space} is likely to be
unstable. However, these developers are not our target audience, we
are trying to focus on developers with a moderate amount of
contribution record who are trying to join a new project, trying to
use a new API, or aiming to get their contributions accepted in a
project. Similarly, rare APIs may not be accurately represented as the corpus
may not have sufficient number of instances of using such API. 

Many potential improvements to the embedding approaches could be
considered. Since our concern was to demonstrate the feasibility of
the approach, we chose an established and computationally efficient
\textit{Doc2Vec} method. With the field of text analysis rapidly
evolving, we expect that future work will develop more accurate
methods that are likely to vary with the task (API/developer/project/PR
prediction), vary with the programming language, or use alternative
embedding techniques. We also expect further work to refine the
parameters of embedding methods as well. Our largest model took more
than three weeks to fit, limiting the ability to run
performance-optimization experiments.

While demonstrating the use of \textit{Skill-Space} based embedding, 
we only compared our results with a random selection of
APIs/developers/projects. A more practical application would be to use 
our method to predict, for example, which APIs a developer will use in
future, and test the prediction accuracy.

Another potential shortcoming of our approach is that it is
not completely resistant to hacking (similar to most other existing
methods of reporting developer expertise) since it is possible to
generate a number of toy projects that use a specific set of APIs to
give an impression that the developer who set up those projects is
skilled with such APIs. However, this is not completely
straightforward either, since it involves the creation of several
toy projects. Further refinements of our method are in progress to
make it more robust. 

While we model a very large corpus of software, it all represents
open source development. The activity of developers in non-public
repositories and non-public software are not captured in this
analysis. Future work is needed to apply our techniques on
proprietary code bases to ascertain if \textit{Skill Spaces} can be
operationalized in the same way or some adaptations are needed to
take into account the differences in the development process.

In terms of external validity, our method can only account for the
developers' expertise, while it is possible that other factors (e.g.
change in job responsibilities) might influence developers when choosing
APIs to use (H1), or which projects to contribute to (H2), which won't 
be captured by our approach.

\section{Future Work}\label{s:future}

Previous sections discussed a variety of promising approaches for
future work to improve the quality of \textit{Skill Space}
representations and to evaluate alternative ways to capture to what
extent a particular change may require/increase API-related skills. We can use the \textit{Skill Space} embeddings of the developers, projects, and APIs together with more efficient machine learning models to further test the applicability of our approach.

More far-reaching extensions of\textit{Skill Space} would be to
include non-technical skills, such as communication and
collaboration skills that are also very important in establishing
trust. We could, potentially, use traces of development activity
related to developers ability to communicate, write high-quality
code, respond to issues, get pull request accepted and other
important skills. This, however, would require a way to evaluate the
quality of the artifacts a developer produces and the quality of the
practices they employ.

A recent paper~\cite{kam19} utilized WoC as a way to estimate the
reputation of a developer using a tool (DRE) that
serves up developer profiles and provides a broad overview of many
facets of a developer's activity, focusing on  both technical and
social aspects. The \textit{Skill
Space} embedding presented in this paper can be used to enhance such developer
profile tools, and
can also provide recommendations for both the
developers (e.g. similar projects that they might consider joining, 
similar developers they might
want to work with in the future, and similar technologies/APIs they
might consider working with etc.) and the project maintainers (e.g. 
potential contributors who might possess relevant skills).

Further application of our approach might include: a) detecting if a developer is actually a bot by analyzing the concentration of their \textit{skill vector} (similar to~\cite{dey2020botdetection,dey2020botse}); b) checking the alignment between \textit{skill vectors} of different developers for identity resolution (similar to~\cite{fry2020idres}); c) analyzing the \textit{skill vectors} of the developers in a project to infer the transparency of the corresponding software supply chain~\cite{dey2019patterns,dey2018dependency,Amreen2019bookchapter,dey2018usageQuality,masupply,Dey2020qualityEMSE}.

\section{Conclusion}\label{s:conclusion}

We have established a proof-of-concept for \textit{Skill Space}: an
approach to represent packages (APIs), developers, languages, and projects in
the same vector space with a topology that satisfies several
practically-relevant criteria, such that the representations of
developers (projects) in \textit{Skill Space} are
similar to the representations of the APIs they use
(contain). Furthermore,  \textit{Skill Space} representations are
predictive of the future API usage by developers, 
developers joining new projects, and it also affects the probability of a
developer's pull requests being accepted. Finally, these representations are
aligned with developers' self-reported expertise.

As with all data-intensive techniques, only entities that have
sufficient data can be accurately represented, but a large volume of
public data from OSS projects can help. The simplicity of the
proposed estimation techniques make it easy to apply them within
enterprises, producing company-specific \textit{Skill Spaces}
that could be integrated with the OSS data.

Two observations were primary motivator for us to
conceptualize the medium-granularity expertise created from the
implicitly defined relationships among APIs in the vast corpus of
open source software projects:
\begin{enumerate}
    \item Contemporary software development increasingly involves
  complex dependency chains with much of the software product
  depending on software developed by unknown and unfamiliar teams;
  \item The ability of developers to use specific libraries and
  frameworks (in the dependency chains noted above) is an important
  factor that determines their ability to complete programming tasks.  
\end{enumerate}

We hope that the progress on measuring and understanding technical
aspects of expertise may prove helpful in developing approaches that
establish trust between maintainers and contributors who had no
prior interactions. We also hope that it may shed some light on the
causes of the vast differences in programmer productivity and help
research on developer learning trajectories. 
We shared
source code and the datasets used in this work, and are also working on making them accessible 
via a web interface through the World of Code website (\url{https://worldofcode.org/}), which can be used to calculate individual vectors and similarities between different entities, with the intention of
facilitating replications, further improvements in the approaches to
construct \textit{Skill Space}, and, more generally, supporting
further studies in this area.

\bibliographystyle{IEEEtran}
\newcommand{\enquote}[1]{``#1''}
\bibliography{all,audris,references}

\end{document}